\documentstyle[aps,epsfig]{revtex}
\begin{document}
\draft
\tighten
\title{Formation Of Quark Matter In Neutron Stars}
\author{S. C. Phatak}
\address{Institute of Physics, Bhubaneswar 751 005, India}
\author{P. K. Sahu}
\address{INFN, Sezione di Catania, 57 Corso Italia, I-95129 Catania, Italy}
\maketitle

\begin{abstract}
At very large densities and/or temperatures a quark-hadron phase
transition is expected to take place. Simulations of QCD on lattice
at zero baryon density indicate that the transition occurs at $T_c
\sim  150-170$ MeV. The calculations indicate that transition is 
likely to be second order or a cross over phenomenon.
Although the lattice simulations have not given any indication on when 
the transition occurs at nonzero baryon density, the transition 
is expected to occur around the densities of few
times nuclear matter density. Also, there is a strong reason to
believe that the quark matter formed after the phase transition is in
colour superconducting phase. The matter densities in the interior of
neutron stars are expected to be several times the nuclear matter density
and therefore the neutron star cores may possibly consist of quark
matter. One then expects that this quark matter is formed during the
collapse of supernova.  
Starting with the assumption that the quark matter, when formed
consists of predominantly u and d quarks, we consider the evolution of
strange quarks by weak interactions in the present work.
The reaction rates and time required to
reach the chemical equilibrium are computed here.
Our calculations show that the chemical equilibrium is reached in
about $10^{-7}$ seconds. Further more during and immediately after the
equilibration process 
enormous amount of energy is released and copious numbers of neutrinos 
are produced. We show that for reasonable models of nuclear equations
of state the amount of energy released 
 could be as high as $10^{53}$ ergs and as many as $10^{58}$
neutrinos may be emitted during the quark matter formation.
%\end{quote}
\end{abstract}


\section{Introduction}

When the nuclear matter is compressed and/or heated, it is expected to 
undergo hadron-quark phase transition with the phase at higher densities 
or temperatures consisting of deconfined quarks and gluons. A schematic
phase diagram for this transition is shown in Fig(1). Although the
idea of quark-hadron phase transition is, by now, widely accepted, the 
details of the transition are not yet completely understood. Recent
lattice-QCD calculations\cite{lat} show that for three flavour matter, 
the transition
temperature at zero baryon chemical potential ( or baryon density ) is 
about 170 MeV. These calculations also show that at the transition
point, the system consists of deconfined but strongly interacting
quarks and gluons and only when the temperature reaches 230 MeV and
above the interactions between quarks and gluons become weak and the
matter can be regarded as  
( almost ) free quarks and gluons. Above this temperature the energy
density and pressure start approaching the value for the gas of
non-interacting relativistic  particles. The lattice calculations
indicate that at the transition temperature the energy density is around
0.6 $GeV/fm^3$ and when the temperature reaches 230 MeV or so the
energy density is around 3 $GeV/fm^3$\cite{lat}. Note that, according
to earlier estimates\cite{trd} of the energy density at the transition
temperature was expected to be higher at 2
$GeV/fm^3$, which is 3 to 4 times larger than the current lattice
estimate. Depending on the mass of the strange quark, the phase transition
may be second order or cross-over phenomenon. It is expected that when the 
baryon chemical potential is increased from zero, at some point a
tricrital point is reached and beyond that point the transition is
expected to be first order. At present, the lattice QCD calculations
are not in position tell much about the chemical potential or baryon
density  at which the transition takes place when the temperature is
small. However, it is reasonable to assume that the energy density at
which the phase transition occurs does not depend sensitively on the 
temperature or chemical potential. In that case, at small temperatures
the baryon density at the transition is expected to be around
3-5 times nuclear matter density\cite{foot1}.

Recent theoretical investigations have\cite{csc} shown that, because of 
attractive quark-quark interaction in isospin-singlet and
colour-$\overline 3$ channel, the quark matter, after the phase
transition, should be in colour-superconducting phase. The Cooper
pairs of the colour-superconducting phase will have diquark structure
with the colour being in $ \overline 3$ channel. Although the
details of the colour-superconducting phase are not yet known it is
expected that the superconducting gap 
should be tens of MeV and therefore the colour superconductivity
should persists at temperatures of tens of MeV. This is likely to have 
interesting consequences on the cooling of neutron stars. In
particular, if if the superconducting gap is tens of MeV, the
elastic scattering of neutrinos from the quarks will be suppressed and 
this will make the quark matter transparent to neutrinos. We shall
come to this point later.

\begin{figure}[h]
\epsfxsize=12.5cm
\centerline{\epsfbox{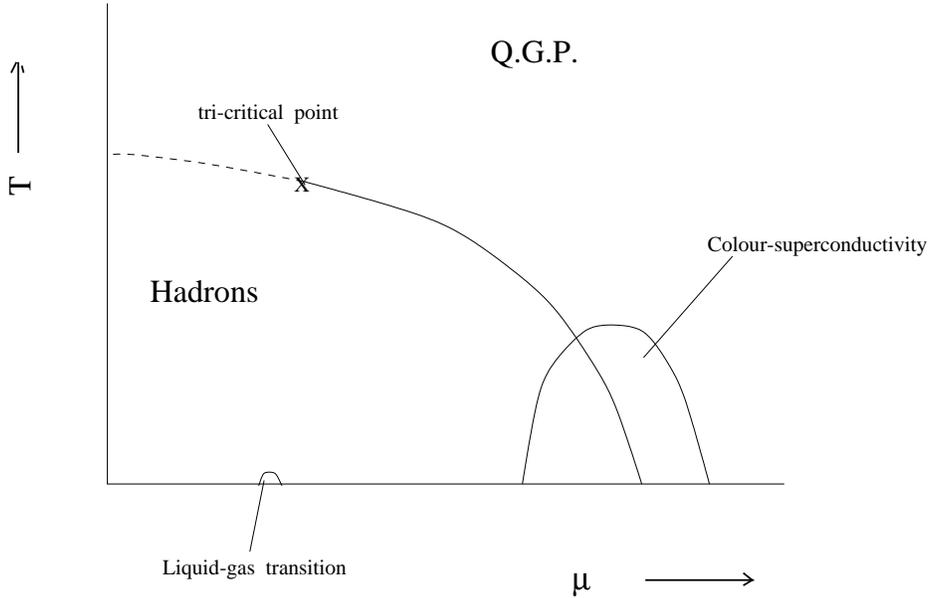}}
\caption{The schematic phase diagram}
\end{figure}

The preceding discussion implies that the 
nuclear matter is likely to undergo hadron-quark phase transition when the
nuclear densities increase above $3-5 \times \rho_{NM}$ where
$\rho_{NM}$ is the equilibrium nuclear matter density ( $0.17 fm^{-3}$ 
). Calculations based on a widely different nuclear equations of
state\cite{ns} show that the baryon number densities in the cores of
neutron stars are quite large ( 7 times $\rho_{NM}$ or larger ) and
therefore it is reasonable to expect that the cores of the neutron
stars would consist of quark matter. In the present work, we shall be
investigating the details of the formation of the quark matter in such
situations. In
particular, we shall be considering the scenario where the baryonic
matter at the phase transition contains few strange particles and
therefore after the transition the quark matter will consist of
predominantly u and d quarks. This is not a chemically equilibrated
configuration of the quark matter. This can be understood as
follows. At the transition point, the matter will predominantly
consist of d and u quarks and some electrons, with d quark density
being roughly two times u quark density and practically no s
quarks. Thus d and u quark chemical potentials will be large ( much
larger than the s quark mass ). As a result, it would be energetically
favourable for the d quarks at the Fermi surface to decay into s quarks
by weak interaction till chemical equilibrium between d and s quarks
is reached. At the chemical equilibrium, the chemical 
potentials of the three quark species will be almost equal. Further
more, since the baryon density of the quark matter is larger than 
3-5 $\times \rho_{NM}$,
the chemical potentials of the quarks will be large in comparison with
the quark masses. This implies that the densities of the quarks of
three flavours will be almost equal. Therefore, the strangeness
fraction ( defined as the ratio of strangeness density to baryon
number density) would be close to unity in the quark matter. On the
other hand, the strangeness fraction is expected to be close to zero
in the hadronic phase\cite{foot2}.
    Thus, the quark matter formation in 
neutron stars should  be associated with the generation of
strangeness. 

In addition to the production of strangeness the equilibration process 
will liberate considerable amount of energy which will result in the
increase in the temperature of the quark matter. The reason for this
is the fact that the weak interaction reactions will be converting the 
d quarks having large energy ( since these are at the Fermi surface )
to s quarks having small energies since the initial Fermi momentum of
d quarks is much larger than the Fermi momentum of s quarks. The
difference in the energies of the d and s quarks will go into  raising
of the temperature the quark matter. In addition, the semi-leptonic
weak interactions, which will be occurring during the chemical
equilibration will produce a large number of neutrinos. Further more,
after the chemical equilibration, the semi-leptonic weak reactions
are expected to continue and contribute to the cooling of the quark
matter by loosing energy to the out-going neutrinos. In the
present work, we have computed the different weak reaction rates which 
are responsible for the chemical equilibration process. Using these
reaction rates the chemical equilibration time, change in the
temperature and pressure of the quark matter and the number of
neutrinos produced during equilibration are calculated. In our
calculations the reaction rates are computed using standard weak
interaction theory and we have assumed that the constituents of the
quark matter are in thermal equilibrium during the chemical equilibration
process. This is a reasonable assumption since the chemical
equilibration precess is due to weak interaction where as the thermal
equilibration is achieved by strong and electromagnetic interactions
which have larger reaction rates.

The paper is organized as follows. In Section II, the reaction rates
for different weak interactions are computed and the evolution
equations for densities of different constituents of the quark matter
are defined. The expression for neutrino emission rates, rate of
change of pressure, internal energy and energy liberation into
neutrinos are also given in Section II. In Section III the results of
the computations are presented and discussed. Finally the implications 
of these results on supernova collapse is discussed in Section IV.

\section{Weak Interaction Rates}

As mentioned in the preceding Section, when the transition from
hadron to quark matter takes place in stellar object ( a neutron star
or a proto-neutron star formed during the collapse of a white dwarf ), 
the quark matter predominantly consists of u and d quarks, apart from
some electrons and gluons if the temperature of the star is
large. Particularly, the s quark density is small when the transition
takes place. Generally, the temperature to be few tens of MeV in 
a proto-neutron star and less than 0.1 MeV for neutron stars at the
transition. As a result, the gluons are not expected to play an
important role. To begin with, we assume the baryon density ( $\rho_B$
) to be same as that before the phase transition\cite{foot3}.
    The u and d quark and electron densities are
determined from baryon density, charge neutrality and
$\beta$-equilibrium ( among u, d quarks and electrons ):
\begin{eqnarray}
\frac{1}{3} ( \rho_d + \rho_u )&  = & \rho_B \nonumber \\ 
\frac{2}{3} \rho_u - \frac{1}{3} \rho_d - \rho_e & = & 0 \nonumber \\
\mu_d & = & \mu_u + \mu_e.
\end{eqnarray}
Here $\mu_i$ is the chemical potential of the particle species $i$ and
the particles are assumed to have Fermi distribution $
F(\epsilon_i,\mu_i,T) = \frac{1}{e^{( \epsilon_i - \mu_i)/T}+1}$ with
$\epsilon_i$ and $\mu_i$ being the energy and chemical potential of
the particle of species i and T the temperature. If the strange
particle density is assumed to be nonzero before the transition, the
initial strange quark density is determined so that the strangeness
fraction after the transition is same as that before the
transition. Then the charge conservation and baryon number density
conservation equations above get modified to include s quarks. 

The chemical equilibration is reached by the weak interactions among
the quarks and electrons. These are:
\begin{eqnarray}
d + u & \leftrightarrow & s + u \\
d ( s ) + e^+  & \rightarrow & u + \overline \nu_e \\
u + e^- & \rightarrow & d ( s ) + \nu_e.
\end{eqnarray}
The first reaction above is nonleptonic weak reaction and it does not
involve neutrino emission. The neutrinos are emitted in other two
reactions. Note that both electrons and positrons are included in the
reactions above since the temperature of the matter is generally
larger than the electron mass and generally the electron chemical
potential is less than the temperature. In addition, one also has
reactions which have three body 
final state:
\begin{eqnarray}
d ( s ) & \rightarrow & u + e^- + \overline \nu_e \\
u & \rightarrow & d ( s ) + e^+ \nu_e.
\end{eqnarray}
It turns out that the reaction rates of the last two reactions are much
smaller. We therefore do not include these reactions in our
calculations. 

Using the standard V-A theory of the weak interactions\cite{weak}, the
transition matrix elements for these reactions can be calculated. One
can then compute the reaction rates by 
integrating the square of the transition matrix element over the momenta 
of the particles involved in the reaction, average over the spins
of incident particles and sum over the spins of product
particles. After some integrations the reaction rates can be written
in the form
\begin{eqnarray}
R_{d \rightarrow s}(t) & = & \frac{18 G^2_F \sin^2 \theta_c \cos^2
  \theta_c }{ ( 2 \pi )^5 }\int dp_d dp_{u_1} dp_{u_2} dp_s p_d
p_{u_1} p_{u_2} p_s \delta (
\epsilon_d + \epsilon_{u_1} -\epsilon_s - \epsilon_{u_2} ) \nonumber \\
 & & F(\epsilon_d, \mu_d, T) F(\epsilon_{u_1}, \mu_u, T) \Big ( 1 -
F(\epsilon_s, \mu_s, T) \Big ) \Big ( 1 - F(\epsilon_{u_2}, \mu_u, T)
\Big )      \nonumber \\
 & & \int_{max\{ |p_d-p_{u_1}|,|p_s-p_{u_2}| \}}^{min\{p_d+p_{u_1} ,
   p_s-p_{u_2} \} } dP \Big ( 1 + \frac{p_d^2 + p_{u_2}^2 -P^2}{2
   \epsilon_d \epsilon_{u_2}} \Big ) \Big ( 1 + \frac { p_s^2 +
   p_{u_1}^2 - P^2 } { 2 \epsilon_s \epsilon_{u_1}} \Big ) \\
R_{s \rightarrow d}(t)  & = &  \frac{18 G^2_F \sin^2 \theta_c \cos^2
  \theta_c }{ ( 2 \pi )^5 }\int dp_d dp_{u_1} dp_{u_2} dp_s p_d
p_{u_1} p_{u_2} p_s \delta (
\epsilon_s + \epsilon_{u_1} -\epsilon_d - \epsilon_{u_2} ) \nonumber \\
 & & F(\epsilon_s, \mu_s, T) F(\epsilon_{u_1}, \mu_u, T) \Big ( 1 -
F(\epsilon_d, \mu_d, T) \Big ) \Big ( 1 - F(\epsilon_{u_2}, \mu_u, T)
\Big )      \nonumber \\
 & & \int_{max\{ |p_s-p_{u_1}|,|p_d-p_{u_2}| \}}^{min\{p_s+p_{u_1} ,
   p_d-p_{u_2} \} } dP \Big ( 1 + \frac{p_s^2 + p_{u_2}^2 -P^2}{2
   \epsilon_s \epsilon_{u_2}} \Big ) \Big ( 1 + \frac { p_d^2 +
   p_{u_1}^2 - P^2 } { 2 \epsilon_d \epsilon_{u_1}} \Big ) \\
R_{d \rightarrow u}(t) & = & \frac{6 G^2_F \cos^2
  \theta_c }{ ( 2 \pi )^5 } \int dp_d dp_{e^+} dp_u dp_\nu p_d
p_{e^+} p_u p_\nu \delta ( \epsilon_d + \epsilon_{e^+} - \epsilon_{u}
- \epsilon_\nu ) \nonumber  \\
& & F(\epsilon_d, \mu_d, T) F(\epsilon_{e^+}, \mu_{e^+}, T) \Big ( 1 -
F(\epsilon_u, \mu_u, T) \Big ) \nonumber \\
 & & \int_{max\{ |p_d-p_{u_1}|,|p_s-p_{u_2}| \}}^{min\{p_d+p_{u_1} ,
   p_s-p_{u_2} \} } dP \Big ( 1 + \frac{p_d^2 + p_{u_2}^2 -P^2}{2
   \epsilon_d \epsilon_{u_2}} \Big ) \Big ( 1 + \frac { p_s^2 +
   p_{u_1}^2 - P^2 } { 2 \epsilon_s \epsilon_{u_1}} \Big ) \\
R_{s \rightarrow u}(t) & = & \frac{6 G^2_F \cos^2
  \theta_c }{ ( 2 \pi )^5 } \int dp_s dp_{e^+} dp_u dp_\nu p_s
p_{e^+} p_u p_\nu \delta ( \epsilon_s + \epsilon_{e^+} - \epsilon_{u}
- \epsilon_\nu ) \nonumber  \\
& & F(\epsilon_s, \mu_s, T) F(\epsilon_{e^+}, \mu_{e^+}, T) \Big ( 1 -
F(\epsilon_u, \mu_u, T) \Big ) \nonumber \\
 & & \int_{max\{ |p_d-p_{u_1}|,|p_s-p_{u_2}| \}}^{min\{p_d+p_{u_1} ,
   p_s-p_{u_2} \} } dP \Big ( 1 + \frac{p_d^2 + p_{u_2}^2 -P^2}{2
   \epsilon_d \epsilon_{u_2}} \Big ) \Big ( 1 + \frac { p_s^2 +
   p_{u_1}^2 - P^2 } { 2 \epsilon_s \epsilon_{u_1}} \Big ) \\
R_{u \rightarrow d}(t) & = & \frac{6 G^2_F \sin^2 \theta_c 
   }{ ( 2 \pi )^5 } \int dp_d dp_{e^+} dp_u dp_\nu p_d
p_{e^+} p_u p_\nu \delta ( \epsilon_u + \epsilon_{e^-} - \epsilon_{d}
- \epsilon_\nu ) \nonumber  \\
& & F(\epsilon_u, \mu_u, T) F(\epsilon_{e^-}, \mu_{e^-}, T) \Big ( 1 -
F(\epsilon_d, \mu_d, T) \Big ) \nonumber \\
 & & \int_{max\{ |p_d-p_{u_1}|,|p_s-p_{u_2}| \}}^{min\{p_d+p_{u_1} ,
   p_s-p_{u_2} \} } dP \Big ( 1 + \frac{p_d^2 + p_{u_2}^2 -P^2}{2
   \epsilon_d \epsilon_{u_2}} \Big ) \Big ( 1 + \frac { p_s^2 +
   p_{u_1}^2 - P^2 } { 2 \epsilon_s \epsilon_{u_1}} \Big ) \\
R_{u \rightarrow s}(t) & = & \frac{6 G^2_F \sin^2 \theta_c 
   }{ ( 2 \pi )^5 }\int dp_s dp_{e^+} dp_u dp_\nu p_s
p_{e^+} p_u p_\nu \delta ( \epsilon_u + \epsilon_{e^-} - \epsilon_{s}
- \epsilon_\nu ) \nonumber  \\
& & F(\epsilon_u, \mu_u, T) F(\epsilon_{e^-}, \mu_{e^-}, T) \Big ( 1 -
F(\epsilon_s, \mu_s, T) \Big ) \nonumber \\
 & & \int_{max\{ |p_d-p_{u_1}|,|p_s-p_{u_2}| \}}^{min\{p_d+p_{u_1} ,
   p_s-p_{u_2} \} } dP \Big ( 1 + \frac{p_d^2 + p_{u_2}^2 -P^2}{2
   \epsilon_d \epsilon_{u_2}} \Big ) \Big ( 1 + \frac { p_s^2 +
   p_{u_1}^2 - P^2 } { 2 \epsilon_s \epsilon_{u_1}} \Big ). 
\end{eqnarray}
Here the Fermi distribution function F
ensures that the particles before the reaction are in the Fermi sea and the
produced particles are above the Fermi sea. The constants $G_F$ and
$\theta_c$ are the weak interaction ( Fermi ) constant and Cabibo
angle respectively. In
the calculation of the reaction rates we have assumed that the
neutrinos produced in the reaction leave the reaction region and
therefore their chemical potential is taken to be zero. The knowledge
of the reaction rates allows us to compute the rate of change of 
densities of various particle species:
\begin{eqnarray}
\frac{d\rho_d}{dt} & = & R_{sd} - R_{ds} + R_{ud} - R_{du} \\
\frac{d\rho_u}{dt} & = & R_{du} - R_{ud} + R_{su} - R_{us} \\
\frac{d\rho_s}{dt} & = & - R_{sd} + R_{ds} + R_{us} - R_{su} \\
\frac{d\rho_{e^-}}{dt} & = & R_{du} - R_{ud} + R_{su} - R_{us}.
\end{eqnarray}
In addition, the internal energy of the system as well as its pressure 
changes as the matter evolves towards chemical equilibrium. There is a 
slight decrease in the pressure and that is because the equilibration
process increases the degrees of freedom ( addition of s quarks!
). The decrease in the internal energy is because the escaping 
neutrinos carry energy. The rate of change of the internal energy is
given by the neutrino emissivities $\epsilon_\nu$ and
$\epsilon_{\overline \nu}$ which are essentially the equations similar 
to the rate equations above with  the neutrino ( antineutrino ) energy 
included in the momentum integrations. In the rate equations above,
$\rho_{e^-}$ implies the difference of electron and positron
densities. It is straight forward to verify that baryon number and
charge conservation is satisfied by these equations. 

Given the reaction rates and the neutrino emissivities at time $t$, one
can compute the number densities of different species and the total
internal energy after a later time interval $t+dt$. The temperature
and chemical potentials of different species at this instant of time
can be determined from the number densities and internal energy
by assuming thermal equilibrium. From these one can compute new rates
at time $t+dt$.
Hence the evolution of the constituents of the quark matter as a
function of time can be obtained numerically. The computation is
carried out till the 
chemical equilibrium is reached, that is when the chemical
potentials of d and s quarks become equal or when the
strangeness fraction ( defined as the ratio of strangeness density to
baryon density ) saturates.  

In addition to the evolution of number densities of different species
one would like to know the number of neutrinos emitted and the energy
carried by the neutrinos during the equilibration process. These are
obtained by integrating the reaction rates and neutrino emissivities:
\begin{eqnarray}
n_\nu & = & \int dt ( R_{ud}(t) + R_{us}(t)) \\
n_{\overline \nu} & = & \int dt ( R_{du}(t) + R_{su}(t)) \\
E_\nu & = & \int dt ( \epsilon_{ud}(t) + \epsilon_{us}(t) ) \\
E_{\overline \nu} & = & \int dt ( \epsilon_{du}(t) + \epsilon_{su}(t) ),
\end{eqnarray}
where $\epsilon(t)$ is the neutrino emissivity computed by including
the neutrino energy in the integration defining the corresponding
reaction rate. Note that after the chemical equilibrium is reached,
the reaction rates for forward and backward reactions are equal
( $R_{ud}(t) = R_{du}(t)$ and  $ R_{us}(t) = R_{su}(t)$ ). Therefore,
although the densities of different quark species does not change
after chemical equilibration, the neutrino emission still
continues. This is essentially thermal 
emission of neutrinos and is actually responsible for the cooling of
the star. We continue our calculation for $10^{-4}$ sec, which we find 
is much larger than  the chemical equilibration time.

\section{Results}

The calculations described in the previous section have been performed 
for a number of initial conditions ( baryon density, temperature and
strangeness content etc). In addition, computations have been
performed for two different models of the quark matter. One is the bag 
model\cite{bag} in which the masses of u, d and s quarks are 0, 0 and
150 MeV respectively. The other is the chiral colour dielectric ( CCD
) model\cite{CCD} where these masses are 100, 100 and 210 MeV. In some 
sense, the bag model represents a situation when one has asymptotic
freedom and therefore the quark masses are the current quark
masses. The quark masses in CCD model are closer to the ( relativistic 
) constituent quark masses. Since the lattice calculations\cite{lat}
indicate that at the phase transition, the interaction between the
quarks is substantial  it is reasonable to expect that the quark
masses just after the phase transition are closer to constituent
masses. Thus, the actual situation is expected to be between the
results of the bag and CCD models. We find that the 
qualitative behavior of the results for different initial conditions
is quite 
similar. We have therefore considered one set of the results of our
calculation for detailed discussion. The behavior of the temperature,
strangeness fraction and reaction rates is displayed in Fig(2). The 
calculation has been done for initial quark matter temperature of 10
MeV and baryon density of $ 3 \times \rho_{NM}$. The strangeness
fraction at the time of quark matter formation is assumed to be
zero. The lower panel of Fig(2) shows the evolution of 
temperature and strangeness fraction and the upper panel shows the 
different reaction rates as a function of time. 

\begin{figure}[h]
\epsfxsize=13.5cm
\centerline{\epsfbox{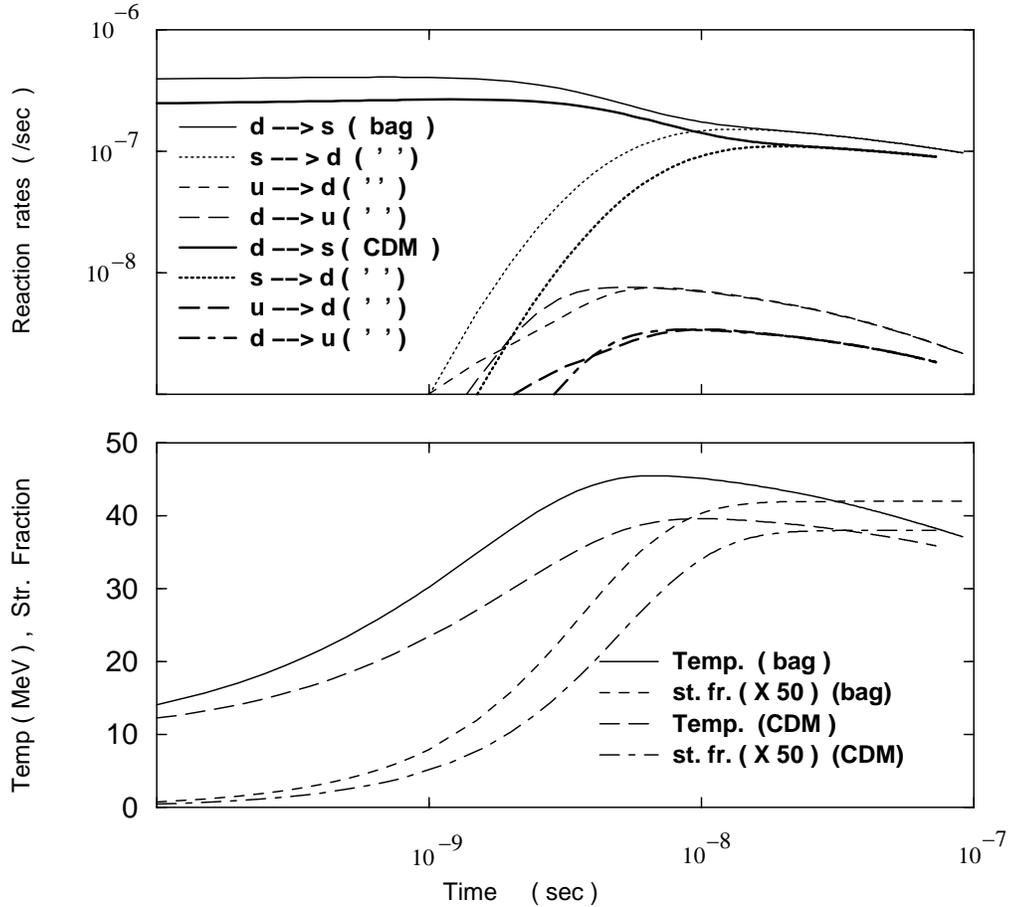}}
\vspace*{0.2cm}
\caption{Results for bag and CCD models. The lower panel displays the
  evolution of temperature and strangeness fraction and the upper
  panel shows the reaction rates}
\end{figure}

The results in Fig(2) show that the chemical equilibrium is reached in
a very short time ( less than $10^{-7}$ seconds ). This time period,
being of the order of typical weak interaction time scale, should not
be surprising. But the important thing is, it is much smaller than the
typical time scales of  
supernova collapse or initial evolution of the proto-neutron
star\cite{proto}. The chemical equilibration time is more or less same 
for the bag and CDM models. Thus, we can conclude that, if the quark
matter is formed in neutron stars, the chemical equilibrium in the
quark matter will be reached very quickly, in a time scale much
shorter than the time scales of the evolution of neutron stars. As
expected, the value of the 
strangeness fraction is close to unity when the chemical equilibrium
is reached. We have not shown the results beyond $10^{-7}$ sec since
the formation of the quark matter because by this time the chemical
equilibrium has already been reached. This one can see from the fact
that the strangeness fraction becomes constant.

\begin{figure}[h]
\epsfxsize=13.5cm
\centerline{\epsfbox{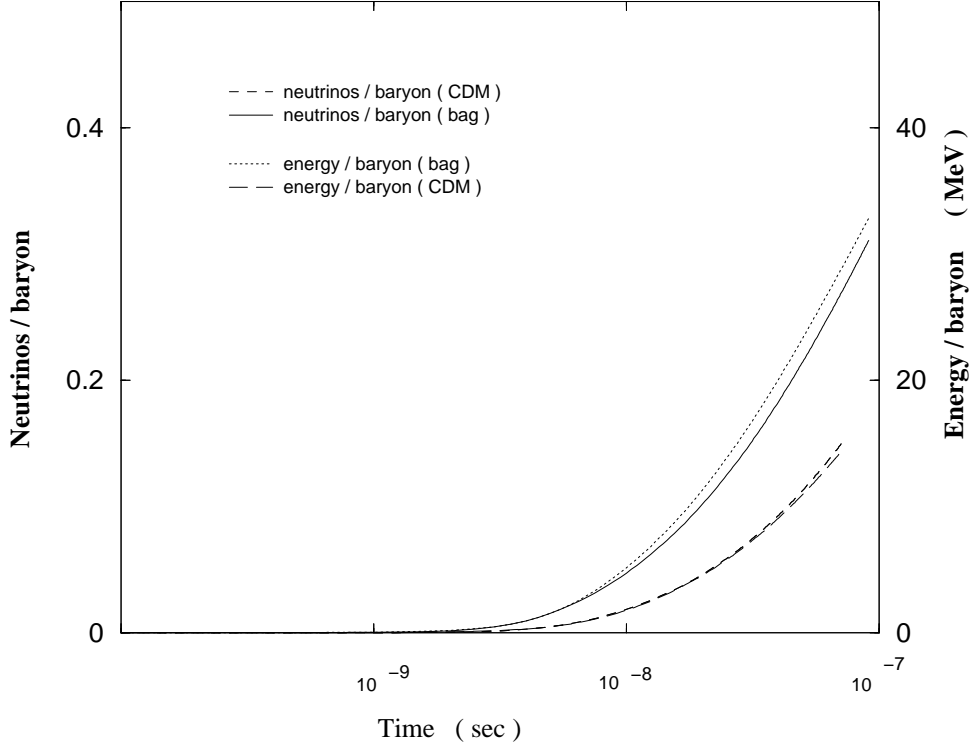}}
\caption{Plot of number of neutrinos/baryon and neutrino energy/baryon 
  emitted during chemical equilibration and after. Note that the
  strangeness fraction saturates around $10^{-8}$ seconds. }
\end{figure}

The calculations show that the temperature of the quark matter rises
sharply during the chemical equilibration process and reaches the peak 
value in about $10^{-8}$ sec. At that stage, the strangeness fraction
has reached about 80 \% of its saturation value. Depending on the
initial temperature and, to some extent on the model, the increase in the
temperature of the quark matter is between 30 and 40 MeV. Thus, the
reactions leading to chemical equilibration of quark matter are
exothermic. As discussed earlier, the 
rise in the temperature is because the d quarks having large momenta
and kinetic energies are converted to s quarks. Since the s quark
chemical potential is small to begin with, their energy is used up in
heating of the quark matter. Note that beyond $10^{-8}$ seconds, the
temperature of the quark matter starts dropping. This decrease is
because the neutrinos emitted in semi-leptonic weak reactions carry
away the energy from the quark matter and as a result there is cooling
of the matter. Neutrino emission is believed to be the most efficient
process  
for the cooling of neutron stars when the temperatures are larger than
1 MeV. We do find that the neutrino emission leads to a rapid cooling
of the quark matter. The calculations, when continued to larger time,
show that the the temperature of the quark matter drops to 10 MeV or
lower in $\sim 10^{-4}$ seconds. 

The top panel of the figure shows different reaction rates. The rate
for $ d + u \rightarrow s + u$ is the largest and clearly dominates
the equilibration process. On the other hand, the reaction rates of
semi-leptonic weak reactions are smaller by more than an order of
magnitude. This means that the chemical equilibrium in the quark
matter is primarily reached by the conversion of d
quarks to s quarks. After  $10^{-8}$ seconds from the formation of
quark matter,
the reaction rates of $d \rightarrow s$ and $s \rightarrow d$
reactions become almost equal. This implies that the d and s quark
chemical potentials have become almost equal but this does not yet
imply chemical equilibration. However, the chemical equilibrium is
soon established. As mentioned earlier, the weak reactions continue
after the chemical
equilibration without altering the composition of the quark
matter. One important point, however,  is that the  
semi-leptonic reactions, which contribute to the production of
neutrinos, continue after equilibration. These reactions gives rise to 
copious production of neutrinos as well as cooling of the matter
because of the loss of the energy from the quark matter to the
produced neutrinos.  

It is clear from the reaction rates that the semi-leptonic reactions
play only a minor role in the process of chemical equilibration as
such. Never the less, these reactions do take place during the chemical
equilibration process and after words. As discussed above, this leads
to a release of energy from the quark matter in the form of neutrinos. In
order to get some idea about the number of neutrinos emitted and the
energy carried by them, we have plotted the
number of neutrinos/baryon and energy/baryon emitted since the start
of the equilibration process in Figure(3). This figure shows that about 0.03
neutrinos/baryon are emitted during the chemical equilibration and
the per baryon energy released in neutrinos is about 3 MeV. That means 
the average energy carried by the neutrinos is of the order of 100
MeV. Thus, although the number of neutrinos is not large, the energy
carried away by these neutrinos is quite substantial. Further, the neutrino
emission continues after equilibration and a lot more neutrinos are
emitted. For example, by $10^{-6}$ seconds since the beginning of the
equilibration process, about 1 neutrino/baryon is emitted and the
average energy carried by these neutrinos is larger than 50
MeV. Since the cooling of the quark matter is immediately followed by
the chemical equilibration, one can argue that during the quark
matter formation and subsequent chemical equilibration large 
number of neutrinos are produced and enormous amount of energy in form of
neutrino kinetic energy is released. Implication of this on dense
stellar objects 
will be discussed in the following Section.

\section{Dense Stellar Objects}

The study of the constitution of the dense stellar objects has
attracted considerable amount of attention recently. Apart from the
fact that these objects ( neutron stars, proto-neutron stars etc )
constitute the test bodies for the theory of general relativity, the
matter in the interior of these objects is at extremely high
density. For example, typical neutron star has mass larger than solar
mass, radius of about 10 km and the density in the interior exceeding
the saturation density of nuclear matter ( $\rho_{NM}$ ). In a broader 
perspective, heavy ion collisions, neutron stars and early universe
correspond to the different regions of the phase diagram of the matter 
at extreme densities and temperatures. For example, in neutron stars
the matter is found at very low density at the surface and very high ( 
5-10 times  $\rho_{NM}$ \cite{Sahu00} ) in the core. In heavy ion
collisions also, 
the nuclear matter is probed at varying densities. The main difference 
between the two is that in neutron stars, the matter is at temperature 
smaller than the baryon chemical potential where as in heavy ion
collisions, the temperature is generally much larger. Thus, the study
of heavy ion collisions and neutron stars is complementary and one may 
use inferences drawn in one field in the investigation of the other
field. 

Long time ago, Baade and Zwicky \cite{Baad34} had suggested that neutron
stars may be formed in supernova explosions. They had proposed that
when the mass of the iron core of a massive star exceeds the
Chandrasekhar limit, it would start collapsing. During the collapse, a
large amount of gravitational energy would be released, which blows 
away the  
mantle of the star and the collapsed core may form a neutron star.
In this mechanism, the maximum mass of neutron stars should exceed 
1.4$M_\odot$.
By now it is understood that the neutron stars are indeed formed after
supernova 
collapse if the mass of pre-supernova star is not too
large. The evolution of the stellar object from
pre-supernova stage to proto-neutron star and later into a neutron
star has been studied in details\cite{evol}. The constitution of
neutron stars has also been studied using different equations of state 
of the nuclear matter\cite{ns,wal,Sahu00}. These studies indicate that the
baryon density in the cores of proto-neutron stars as well as neutron
stars could be several times $\rho_{NM}$.  This implies that the cores
of these objects  are likely to consist of exotic states of nuclear
matter such as strange matter, kaon condensate, quark matter etc. 
In this paper we are considering the last possibility, namely the
formation of quark matter. As discussed in the Introduction, the
lattice calculations\cite{lat} as well as the recent experiments at
CERN\cite{expt} seem to indicate that the quark matter is likely to be
formed when the energy density of the hot, low baryon density 
matter reaches 600 $MeV/fm^3$ or more. We extrapolate this result to
cold and dense nuclear matter and assume that the quark matter is
formed when the energy density reaches 600 $MeV/fm^3$. This energy
density is reached when the baryon number density is between 0.5 and
0.8  $fm^{-3}$ which  corresponds to 3-5 $\times \rho_{NM}$. 

An estimate of the amount of matter in the neutron star that is
converted to quark matter can be obtained as follows. The properties
of neutron stars without rotation are computed by solving the
Tolman-Oppenheimer-Volkoff (TOV) equation\cite{Hart67,Misn70},which
is essentially an equation for hydrostatic equilibrium in the
star. The solution of TOV equation requires the equation of state of
the matter. Using a variety of nuclear equations of state\cite{ns,wal} 
the TOV equation has been solved and the properties of neutron stars
computed. One of the result of these 
calculations is the variation of nuclear density (density profile) and 
star mass (mass profile) as a function of the distance from the center of 
the star for a given initial value of central density. In the example
considered here, we choose the central density of seven times nuclear
matter density. In Fig(4) density and mass profile
for a soft\cite{Prak88} (incompressibility, K=120 MeV) and a 
hard\cite{Wiri88} (incompressibility, K=224 MeV) equations of state
has been plotted.  

\begin{figure}[h]
\epsfxsize=13.5cm
\centerline{\epsfbox{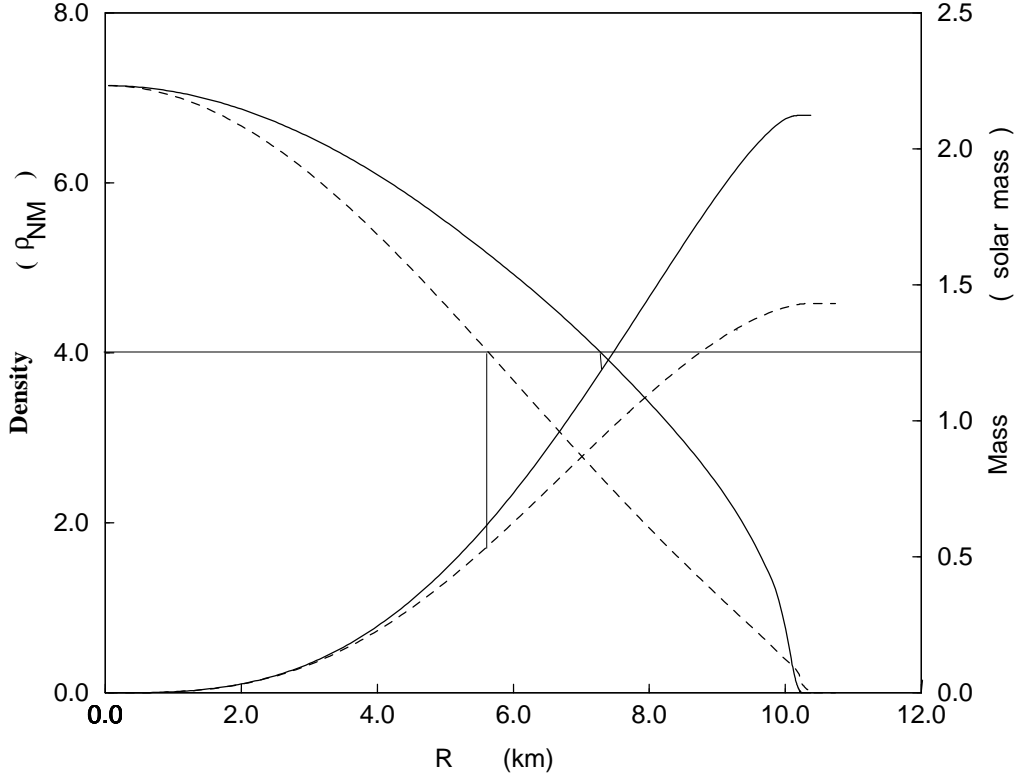}}
\vspace*{0.2cm}
\caption{ Plot of the baryon number density ( scale on the left ) and
  mass contained within distance r from the center of the neutron star
  ( scale on the right ). The baryon number 
  density is in units of $fm^{-3}$, the mass in units of solar mass
  and the distance is in units of kilometers. The horizontal line is
  drawn at 4 times nuclear matter density and the amount of nuclear
  matter having density larger than this value can be read off from
  the graph. The two curves are for a stiff and a soft equation of
  state. } 
\end{figure}

Assuming that the phase transition to quark matter takes place when
the matter density increases above $ 0.7 fm^{-3}$, which is about $4
\times \rho_{NM}$, we have drawn a horizontal line at $4 \times
\rho_{NM}$. This line intersects baryon density profile curves for
soft and hard equations of state at ( about ) 6 and 8 kilometers
respectively. The matter inside this radius undergoes transition
to,quark matter and therefore the total amount of matter in quark
matter phase is about 0.5 and 1.2 solar masses respectively. We have
checked this estimate for a number of nuclear matter equations of
state and different central densities. We find that generally the
estimate of the amount of quark matter for all these calculations is
in this range. Thus
it is reasonable to assume that, if the quark matter is formed in the
core of the neutron stars, its amount is expected to be larger than
0.5 solar mass. We shall choose this number for the further
discussion.    

As discussed in the preceding sections, the hadron-quark phase
transition and subsequent chemical equilibration is a rapid process (
time scale of $\sim 10^{-7}$ sec or so ) and therefore the quark matter
would be formed immediately after the formation of the neutron
star. One of the consequences of the transition is that the
temperature of the quark matter increases by 30-40 MeV during
equilibration. The matter then cools very quickly ( $\sim 10^{-4}$ sec 
) by emitting neutrinos. During this process about 1 neutrino per
baryon is emitted and the average energy carried by the neutrinos is
more than 50 MeV. All this occurs within a millisecond from the time
when the phase transition takes place. Assuming that the mass content
of the quark matter is 0.5 solar mass, the total number of baryons in
the quark matter is about $10^{57}$. This means that roughly $10^{57}$ 
neutrinos will be liberated very soon after the phase transition. The
energy carried by these neutrinos will be more than $5 \times 10^{58}$
MeV or $8 \times 10^{52}$ ergs. 

The neutrinos produced soon after the hadron-quark transition in the
core of the star diffuse out to the surface. During this process their 
numbers are expected to
increase\cite{Bah} ( possibly ) by a factor of 10 during their
propagation from the core of the star to its surface\cite{foot4}.
    Thus the
conservative estimate of the number of neutrinos percolating out from
the star would be $\sim 10^{58}$ or larger. Note that this number is
of the order of the neutrinos emitted by standard nuclear
$\beta$ decay processes during the collapse of the white
dwarf\cite{shap83}. In other words, if the quark-hadron phase transition
takes place in a neutron star,  the number of neutrinos emitted will
be larger by a factor of 2 or more.

One of the mechanisms for the supernova explosion is the reheating of
the mantle of the star by out-streaming neutrinos which gives an extra 
outward push to the shock wave which is stalled. If the quark matter
formation takes place immediately following the supernova collapse,
the neutrinos produced during the quark matter formation will add to
the neutrinos produced during the collapse and this would add to the
reheating of the mantle and therefore facilitate the supernova
explosion. Further more, 
the amount of energy carried by these neutrinos is quite 
large, since on the average, these neutrinos carry 50 MeV of
energy. A crude estimate indicates that this energy will be about
$8 \times 10^{52}$ ergs or more, which is, again, comparable to the
energy emitted 
during the supernova collapse. Actually, it is not surprising that
such a huge amount of energy is liberated during quark matter
formation. In the standard supernova mechanism, the energy release is
primarily due to nuclear reactions and the gravitational energy.  This
energy release is expected to be few tens of MeV/nucleon, a typical
energy scale of nuclear physics. In case of quark matter formation,
the energy released is  
related to the difference in chemical potentials of d and s quarks. At 
the beginning of the quark matter formation, this difference is few
hundreds of MeV. As a result, the neutrinos produced during the chemical
equilibration of the quark matter carry such a large amount of
energy. 

At this point, one may note that there is a possibility of formation
of colour superconductivity in the quark matter if the temperature of
the core is sufficiently small\cite{csc}. Estimates indicate that the
superconducting gap is expected to be $\sim$ 50 MeV and the
colour superconducting temperature is of the same order. In that case, 
the interaction of neutrinos with the quarks will be suppressed as
that would require breaking of the colour-superconducting pair. This
would mean that the quark matter would be transparent to the neutrinos 
and as a result, the neutrinos produced during strangeness production
and subsequent cooling will be able to leave the quark core in a short 
time. The details of the propagation of neutrinos in such media need
to be investigated in details. 

It is suggested that the recent discovery of afterglow in Gamma Ray Bursters 
(GRB) occur at very high redshifts ($z\ge$ 1) and these imply a huge amount of 
energy release of $\sim 10^{53}$ ergs.
For example, GRB 971214\cite{Kulk98} and GRB 990123 \cite{Kulk99} have the 
measured redshifts of  3.42 and
1.6 with a energy output of $3 \times 10^{53}$ erg and $3.4\times 10^{54}$
ergs respectively.
Such enormous energies can be produced by different mechanisms.
One such possible mechanism of producing huge energies is the merger of 
two neutron stars,  eventually forming black hole.
The rate of neutron star merger can be calculated from abundance of binary 
pulsars, which is compatible with the rate of GRB of approximately one per
day.
Another possible mechanism  for GRB is the conversion of a neutron
star to a strange star.
Recently, Bombaci and Datta \cite{Bomb00} proposed that the total amount of 
energy liberated in the conversion is in the range of 1-4 $\times 10^{53}$ erg, 
which is in agreement with GRB sources at cosmological distances.
In the conversion of ud- matter to uds- matter, in our calculation the
chemical equilibrium is reached around $10^{-7}$ seconds.
During this process, enormous amount of energy in the order of 100 MeV per 
baryon \cite{Ghos96} is released and copious numbers of neutrinos are produced.
The energy is released and carried away mainly by the neutrinos and this energy
will be about $10^{53}$ erg and more, which is again comparable to
GRB. If, by some mechanism, the neutrino energy is converted into
$\gamma$ ray energy, one may be able to explain the GRB's in terms of
possible hadron-quark phase transition occurring during the formation
of neutron stars. 
However, a detailed discussion on this is beyond the scope of this paper.

\section{Conclusions}

To conclude, we have argued that the nuclear matter having densities
larger than 3-5 $\times \rho_{NM}$ is likely to undergo phase
transition from nuclear to quark matter. We have shown that 
after hadron-quark phase transition, the quark matter is not 
in chemical equilibrium in the beginning. The process of chemical
equilibration, which  
is achieved by weak interactions, leads to production of s quarks. The
dominant reaction for the equilibration is the nonleptonic weak
process which converts d quarks into s quarks, although semi-leptonic
weak processes also occur during and after chemical equilibration.
The equilibration reaction is exothermic and therefore the temperature
of the quark matter rises by 30-40 MeV during the equilibration. A
large numbers of neutrinos are produced during and soon after the
chemical equilibrium is reached. These
neutrinos carry  substantial amount of energy and therefore they are
responsible for the  cooling of the quark matter. 
We argue that, since the nuclear densities  in neutron stars are several
times $\rho_{NM}$, it is very much likely that the quark matter is formed in
the cores of these stars. The quark matter is expected to be formed
during the supernova collapse. We then conclude that the large number
of neutrinos that are produced during and 
after the formation of quark matter in these cores will add to the
neutrinos that are produced during the supernova collapse. These
neutrinos may add to the
reheating of the mantle of the star and help in the supernova
explosion.


\begin{thebibliography}{99}
\bibitem{foot1} The energy density of
  normal nuclear matter is about $ 0.17 \times 938  \sim 150
  MeV/fm^3$. If we ignore the interaction energy, at four times
  nuclear matter density the energy density would be $\sim 600
  MeV/fm^3$, which is the transition energy density indicated by the
  lattice-QCD calculations.
\bibitem{foot2} Actually the 
    strangeness fraction could be nonzero in the hadronic phase. That
    is because the baryonic matter may have strange baryons or may
    have kaon condensate. The calculations based on Walecka
    model\cite{wal} show that one may have a strangeness fraction of
    0.2 at 3-5 $\times \rho_{NM}$. Similarly, kaon
    condensate\cite{kaon}, if formed, may generate strangeness
    fraction of similar order.
\bibitem{foot3} In fact, we
    treat it as an initial condition and perform the computation for a
    set of values of 
    $\rho_B$. Extrapolation of lattice QCD results implies that the
    transition would be occurring at $\rho_B$ of 3-5 times
    $\rho_{NM}$.
\bibitem{foot4} Neutrino 
    transport in neutron stars is still an open subject. However it is 
    estimated\cite{Bah} that the neutrinos produced in nuclear
    interactions ( which are having energies between 10 and 100 MeV )
    degrade their energy and generate a number of secondary neutrinos
    of all species during propagation through the matter. This estimate
    indicates that the number of neutrinos is increased by a factor
    of 10 during the propagation.  The neutrinos produced in quark matter 
    have even higher
    energies so it would be reasonable to expect that in this case the
    increase in the number of neutrinos could be by a factor of 10 or more.
\end{thebibliography}
\end{document}